\documentclass[preprint,showpacs,preprintnumbers,amsmath,amssymb,nofootinbib]{revtex4}

\usepackage{booktabs}
\usepackage{mathrsfs}
\usepackage{epsfig}
\usepackage{graphicx}
\usepackage{dcolumn}
\usepackage{bm}
\usepackage{amsmath}
\usepackage{slashed}
\usepackage{multirow}
\usepackage{subfigure}
\usepackage{epstopdf}
\usepackage{color}

\let\jnfont=\rm
\def\NPB#1,{{\jnfont Nucl.\ Phys.\ B }{\bf #1},}
\def\PLB#1,{{\jnfont Phys.\ Lett.\ B }{\bf #1},}
\def\EPJC#1,{{\jnfont Eur.\ Phys.\ Jour.\ C }{\bf #1},}
\def\PRD#1,{{\jnfont Phys.\ Rev.\ D }{\bf #1},}
\def\PRL#1,{{\jnfont Phys.\ Rev.\ Lett.\ }{\bf #1},}
\def\MPLA#1,{{\jnfont Mod.\ Phys.\ Lett.\ A }{\bf #1},}
\def\JPG#1,{{\jnfont J.\ Phys.\ G}{\bf #1},}
\def\CTP#1,{{\jnfont Commun.\ Theor.\ Phys.\ }{\bf #1},}
\def\ZPC#1,{{\jnfont Z.\ Phys.\ C }{\bf #1},}
\def\JHEP#1,{{\jnfont JHEP \ }{\bf #1},}
\def\Rv{\not{\hbox{\kern-1pt $R$}}}
\def\p{\not{\hbox{\kern-3pt $p$}}}

\newcommand{\bea}{\begin{eqnarray}}
\newcommand{\eea}{\end{eqnarray}}

\newcommand{\bcen}{\begin{center}}
\newcommand{\ecen}{\end{center}}

\newcommand{\ee}{e^+e^-}

\newcommand{\nn}{\noindent}

\newcommand{\beq}{\begin{eqnarray}}
\newcommand{\eeq}{\end{eqnarray}}

\def\t1{\tilde{t_1}}

\def\be{\begin{equation}}
\def\ee{\end{equation}}
\def\bea{\begin{array}}
\def\eea{\end{array}}
\def\beqa{\begin{eqnarray}}
\def\eeqa{\end{eqnarray}}
\def\beqas{\begin{eqnarray*}}
\def\eeqas{\end{eqnarray*}}

\def\bp{\begin{picture}}
\def\ep{\end{picture}}
\def\bc{\begin{center}}
\def\ec{\end{center}}
\def\bfig{\begin{figure}}
\def\efig{\end{figure}}

\def\bit{\begin{itemize}}
\def\eit{\end{itemize}}
\def\nn{\nonumber}
\def\f{\frac}

\def\[{\left[}
\def\]{\right]}
\def\({\left(}
\def\){\right)}

\def\..{\left.}
\def\.{\right.}
\def\tl{\tilde}

\def\la{\leftarrow}

\def\tm{\times}

\def\da{\dagger}

\def\la{\lambda}

\def\ep{\epsilon}

\def\pa{\partial}
\def\pr{\prime}

\preprint{ TU-1053 }
\begin{document}

\title{Simplified TeV leptophilic dark matter in light of DAMPE data}

\author{ Guang Hua Duan$^{1,2,3}$}
\author{ Lei Feng$^{4}$}
\author{ Fei Wang$^{5}$}
\author{ Lei Wu$^{1}$}
\author{ Jin Min Yang$^{2,3,6}$}
\author{ Rui Zheng$^{7}$
\vspace*{.5cm}}

\affiliation{
$^1$ Department of Physics and Institute of Theoretical Physics, Nanjing Normal University, Nanjing, Jiangsu 210023, China\\
$^2$ CAS Key Laboratory of Theoretical Physics, Institute of Theoretical Physics, Chinese Academy of Sciences, Beijing 100190, China\\
$^3$ School of Physical Sciences, University of Chinese Academy of Sciences, Beijing 100049, China\\
$^4$ Key Laboratory of Dark Matter and Space Astronomy, Purple Mountain Observatory, Chinese Academy of Sciences, Nanjing 210008, China\\
$^5$ School of Physics, Zhengzhou University, Zhengzhou 450000, P. R. China\\
$^6$ Department of Physics, Tohoku University, Sendai 980-8578, Japan\\
$^7$ Department of Physics, University of California, Davis, CA 95616, USA
}%

\begin{abstract}
Using a simplified framework, we attempt to explain the recent DAMPE cosmic $e^+ + e^-$ flux excess by leptophilic Dirac fermion dark matter (LDM). The scalar ($\Phi_0$) and vector ($\Phi_1$) mediator fields connecting LDM and Standard Model particles are discussed. We find that the couplings $P \otimes S$, $P \otimes P$, $V \otimes A$ and $V \otimes V$ can produce the right bump in $e^+ + e^-$ flux for a DM mass around 1.5 TeV with a natural thermal annihilation cross-section $<\sigma v> \sim 3 \times 10^{-26} cm^3/s$ today. Among them, $V \otimes V$ coupling is tightly constrained by PandaX-II data (although LDM-nucleus scattering appears at one-loop level) and the surviving samples appear in the resonant region, $m_{\Phi_1} \simeq 2m_{\chi}$. We also study the related collider signatures, such as dilepton production $pp \to \Phi_1 \to \ell^+\ell^-$, and muon $g-2$ anomaly. Finally, we present a possible $U(1)_X$ realization for such leptophilic dark matter.
\end{abstract}
\maketitle

\section{introduction}
The existence of cold dark matter (CDM) has been confirmed by astrophysical experiments, which
provides a natural way to account for many properties of galaxies on large scales. However,
the nature of CDM has remained elusive. Among various hypotheses for CDM, the paradigm of weakly interacting massive particles (WIMPs) is one of the most attractive candidates. So far, the WIMP dark matter has undergone very close and effective experimental scrutiny, such as direct detections by measuring the nuclear recoil imparted by the scattering of a DM and collider searches for mono-$X$ signatures.

Besides these, indirect detections via observing high energy gamma-rays, cosmic-rays and neutrinos may also shed light on the properties of DM. In past years, several DM satellite experiments, such as AMS-02, PAMELA, HEAT and Fermi-LAT, have been launched and reported some intriguing DM evidences. The DArk Matter Particle Explorer (DAMPE) is a new cosmic ray detector~\cite{dampe1,dampe2}, which has great energy resolution (better than $\rm 1.5\% @TeV$ for electrons and gamma rays) and good hadron rejection power (higher than $10^5$). Very recently, DAMPE released their first results about cosmic-ray $e^+ + e^-$ flux up to 5 TeV \cite{dampe-data}. A sharp peak at $\sim 1.4 $ TeV was reported in DAMPE data, which
implies the existence of a nearby monoenergetic electron sources because of the cooling process of
high energy cosmic-ray electrons~\cite{yuan2017,bi}. No associated excess in the anti-proton flux has
been observed. Both astrophysical sources (e.g., pulsars) and DM interpretations are discussed in
Ref.~\cite{yuan2017}. It is found that DM should annihilate to $e^\pm$ or $\{e^\pm,\mu^\pm,\tau^\pm\}$
with 1:1:1 and the mass of DM particle is about 1.5 TeV if the nearby DM sub-halo located at
$\rm 0.1 \sim 0.3$ kpc away from the solar system~\cite{yuan2017}. Several leptophilic DM model have been proposed to explain this excess~\cite{Fan:2017sor,Gu:2017gle}.


In this work, we attempt to explain this tentative cosmic-ray eletron+positron excess by using a
simplified framework, in which the DM sector has no direct couplings to quarks, only couples with
leptons mediated by a scalar or vector field. Such a leptophilic DM can satisfy the measured relic
density at tree level and may accommodate the null results from direct detections by inducing
interactions between dark matter and quarks at the loop level. Many studies have been devoted into
the idea that DM does not interact with quarks at the tree level. Most of these analyses assume
an interaction between DM and leptons to be flavor blind~\cite{ldm-1,ldm-2,ldm-3,ldm-4,ldm-5,ldm-6,ldm-7,ldm-8,ldm-9,ldm-10,ldm-11,ldm-12,ldm-13,ldm-14,ldm-15,ldm-16,ldm-17,ldm-18,ldm-19,ldm-20,
ldm-21,ldm-22,ldm-23,ldm-24,ldm-25,ldm-26,ldm-27,ldm-28,ldm-29,ldm-30,ldm-31,ldm-32,ldm-33}, while a few other studies assume gauged flavor
interactions~\cite{gldm-1,gldm-2,gldm-3,gldm-4,gldm-5,gldm-6,gldm-7,gldm-8,gldm-9,gldm-10}. The leptophilic DM
framework allows for a more general analysis of interactions that involve only DM and leptons at the tree level. It permits different coupling strengths between lepton flavors, off-diagonal flavor couplings, and lepton-flavor violation \footnote{For a review of flavored dark matter, see Ref.~\cite{review-ldm} and the references therein.}.

The structure of this paper is organized as follows. In Section~\ref{section2}, we introduce the effective lagrangian for leptophilic DM and loop induced LDM-hadron interactions. In Section~\ref{section3}, we present our numerical results for the DAMPE excess and discuss the related collider signatures. In Section~\ref{section4}, we give a possible realization of leptophilic DM in $U(1)$ extensions. Finally, we draw our conclusions in Section \ref{section5}.

\section{Simplified Leptophilic Dark Matter}\label{section2}
The main goal of our study is a model independent analysis of leptophilic Dirac fermion DM ($\chi$)
for the DAMPE excess. We parameterize the relevant DM--lepton interactions as
 \begin{eqnarray}
    {\cal L} &\ni&  \Phi_i \bar{\chi}\Gamma_\chi\chi +\Phi_i \bar{\ell}\Gamma_\ell\ell ,~
\end{eqnarray}
where $\Phi_i$ is a mediator field with $i=0,1$ corresponding to spin-0 and spin-1 boson respectively.
We assume that $\Phi_i$ only couples with leptons $e,\mu,\tau$ in our calculations. Then, the Lorentz
structures of $\Gamma_{\chi,\ell}$ are scalar (S), pseudo-scalar (P), vector (V) and axial-vector (A)
interactions given by
\be\label{eq:types}
\begin{array}{l@{\qquad}l@{\qquad}l}
  \text{scalar-type:} &
  \Gamma_\chi = g_\chi^S + ig_\chi^P\gamma_5, &
  \Gamma_\ell = g_\ell^S + ig_\ell^P\gamma_5, \\
  \text{vector-type:} &
  \Gamma_\chi^\mu = (g_\chi^V + g_\chi^A\gamma_5)\gamma^\mu, &
  \Gamma_{\ell\mu} = (g_\ell^V + g_\ell^A\gamma_5)\gamma_\mu ,\\
%
\end{array}
\ee
where $g_\chi$ and $g_\ell$ are the coupling strengthes of the mediator to DM and SM leptons,
respectively.

\begin{figure}[ht]
  \centering
   \includegraphics[width=4in]{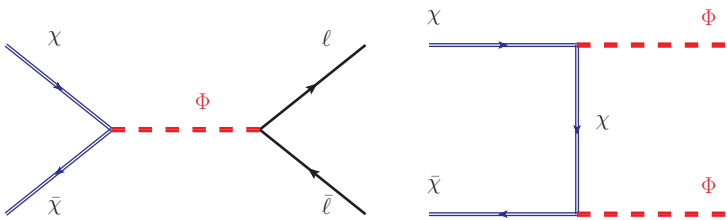}
 \caption{Feynman diagrams for LDM annihilation.}
\label{annihilation}
\end{figure}

In our framework, the dominant LDM annihilation channels are
\begin{equation}
  \chi \bar{\chi} \to \ell\bar{\ell}, \Phi\Phi
\end{equation}
with the corresponding Feynman diagrams in Fig.~\ref{annihilation}. For a pair of LDM, the CP value of the system is given by $(-1)^{S+1}$. Due to the CP and total angular momentum conservation, the quantum states of $|\bar{\chi}\chi\rangle$ are $^3P$ and $^1S$ for the scalar and pseudoscalar mediators, while the corresponding states for vector and axial-vector mediators are $^3S$ and $^1P$, respectively. Then, one can estimate the dominant contributions of LDM annihilation cross section, as shown in Table~\ref{tab:Xsec}. It should be noted that the coupling $A\otimes A$ can produce the $s$-wave contribution, however, which is highly suppressed by mass ratio $m^2_\ell/m^2_\Phi$.

\begin{table}[h]
\begin{tabular}{c@{\quad}|@{\quad}c@{\qquad}cc}
\hline\hline
$\Gamma_\chi \otimes \Gamma_\ell$&  $\sigma v(\chi \chi \to \ell \bar{\ell})$&
\multicolumn{2}{c}{$\sigma(\chi N \to \chi N)/(\frac{\alpha_{em}Z}{\pi m_{\Phi_i}^2})^2$
} \\
\hline
$S\otimes S$ & $p$-wave &  $\alpha_{\rm em}^2 $ & [2-loop]\\
$S\otimes P$ & $p$-wave &  $-$ & \\
$P\otimes S$ & $s$-wave &  $\alpha_{\rm em}^2 v^2$ & [2-loop]  \\
$P\otimes P$ & $s$-wave & $-$ &   \\
$V\otimes V$ & $s$-wave &  $1$ & [1-loop]\\
$V\otimes A$ & $s$-wave & $-$ &  \\
$A\otimes V$ & $p$-wave & $v^2$ &[1-loop] \\
$A\otimes A$ & $p$-wave & $-$ &   \\
\hline\hline
\end{tabular}
\caption{Dominant contribution to LDM annihilation cross section and scattering cross section suppression by small parameters for loop
induced DM-nucleon scattering for eight Lorentz structures. Here $v$ is the DM velocity.}
\label{tab:Xsec}
\end{table}

\begin{figure}[ht]
  \centering
   \includegraphics[width=6in]{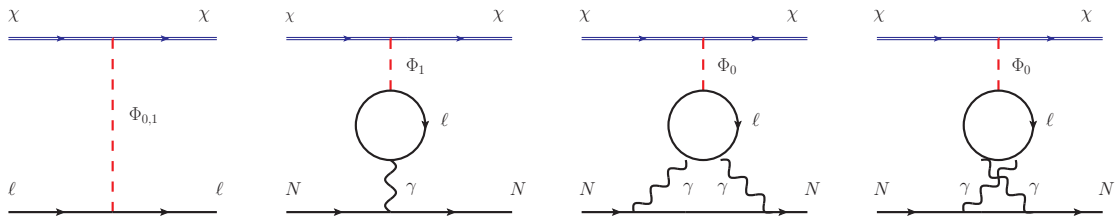}
 \caption{Feynman diagrams for LDM scattering with electron and nucleus.}
\label{dd}
\end{figure}
Since the LDM only interacts with leptons, it can produce the signal by scattering with electron of atom at tree level or with nucleus at loop level in DM direct detection experiments, as shown in Fig.~\ref{dd}. The velocity of DM particles near the Earth is of the same order as the orbital velocity of the Sun, $ v\sim 0.001c$. So the recoil momenta is of order a few MeV, which is much smaller than our mediator mass. Then, we can integrate out heavy mediator fields and obtain the effective operators:
\be\label{eq:4fermi}
{\cal L}_{\rm eff}= \frac{1}{\Lambda^2} \, (\bar\chi \Gamma_\chi \chi) \, (\bar\ell \Gamma_\ell \ell)\,,
\ee
where $\Lambda=m_\Phi/\sqrt{g_\chi g_\ell}$ is the cut-off scale for the effective field theory description. With this setup, one can calculate DM-electron scattering cross section at tree level:
\begin{eqnarray}
\sigma^{\Phi_0}_{\chi e} &=& \frac{m^2_e g^2_\chi g^2_\ell}{m_{\Phi_0}^4}
\left\{
(g_\chi^S g_e^S)^2 +
\left[ (g_\chi^S g_e^P)^2 + (g_\chi^P g_e^S)^2 \frac{m_e^2}{m_\chi^2} \right]
\frac{v^2}{2} +
\frac{(g_\chi^P g_e^P)^2}{3}\frac{m_e^2}{m_\chi^2} v^4 \right\},\\
\sigma^{\Phi_1}_{\chi e} &=& \frac{m^2_e g^2_\chi g^2_\ell}{m_{\Phi_1}^4}
\left\{ (g_\chi^V g_e^V)^2 + 3(g_\chi^A g_e^A)^2
+\left[ (g_\chi^V g_e^A)^2 + 3(g_\chi^A g_e^V)^2 \right]
\frac{v^2}{2}\right\} ,
\label{dm-e}
\end{eqnarray}
We can find that DM-electron scattering cross sections for $S\otimes P$, $P\otimes S$ and $P\otimes P$ couplings are suppressed by both small mass ratio $m_e/m_\chi$ and low velocity $v \sim 10^{-3}$, while for $V\otimes A$ and $A \otimes V$ couplings the cross sections are only suppressed by velocity. All of them are below the current sensitivity of DM-electron scattering experiments.

The loop induced DM-nucleus scattering cross sections for spin-1/0 mediator at one-loop/two-loop level in leading log approximation~\cite{Kopp:2009et} are given by:
\begin{eqnarray}
\sigma^{\Phi_0}_{\chi N} &=& \frac{\mu_N^2}{\pi} (\frac{\alpha_{em}Z}{\pi m_{\Phi_0}^2})^2(\frac{\alpha_{em}Z}{\pi })^2(\frac{\pi^2}{12})^2(\frac{\mu_N v}{m_\ell})^2[2(g_\chi^Sg_\ell^S)^2+\frac{4}{3}(g_\chi^Pg_\ell^S)^2 v^2\frac{\mu_N^2}{m_N^2}]\label{dm-n-1}\\
\sigma^{\Phi_1}_{\chi N} &=& \frac{\mu_N^2}{9\pi} (\frac{\alpha_{em}Z}{\pi m_{\Phi_1}^2})^2[\sum_{\ell=e,\mu,\tau}{\rm log}(\frac{m_\ell^2}{\mu^2})]^2[(g_\chi^Vg_\ell^V)^2+(g_\chi^Ag_\ell^V)^2 v^2(1+\frac{1}{2}\frac{\mu_N^2}{m_N^2})]\label{dm-n-2}
\end{eqnarray}
where $m_N$ and $Z$ are the nucleus's mass and charge respectively, and $\mu_N=\frac{m_\chi m_N}{m_\chi+m_N}$ is the reduced mass of DM-nucleus system. The above two-loop result of $\sigma^{\Phi_1}_{\chi N}$ is obtained by using operator product expansion in heavy lepton approximation. We set the renormalization scale $\mu=m_{\Phi}$ and both nuclear form factors $F(q)$ for $\Phi_1$ and $\tilde{F}(q)$ for $\Phi_0$ to unity for simplicity. According to Eq.~\ref{dm-n-1} and~\ref{dm-n-2}, we present the scattering cross section suppression by small parameters for loop induced DM-nucleon scattering for eight Lorentz structures in Table.~\ref{tab:Xsec}. It can be seen that the DM-nucleus scattering cross sections for $P\otimes S$ and $A\otimes V$ couplings are suppressed by $v^2$, as comparison with $S\otimes S$ and $V\otimes V$ couplings.

\section{Numerical results and discussions}\label{section3}
\begin{figure}[ht]
  \centering
  \includegraphics[width=4in]{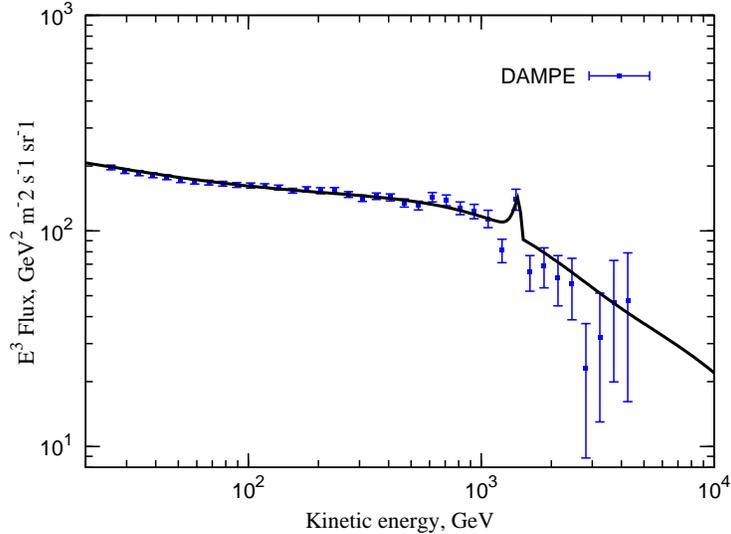}
\caption{Total $e^+ + e^-$ flux of 1.5 TeV DM that annihilates into leptons with the branching ratio $e:\mu:\tau=1:1:1$ for fitting AMS-02 AND DAMPE data. The mass of nearby subhalo is assumed as $1 \rm \times 10^{8} m_\odot$ with a distance 0.1 kpc away from the solar system.}
\label{dampe}
\end{figure}
According to the analysis of Ref.~\cite{yuan2017}, the excess of $e^+ + e^-$ flux in DAMPE can be interpreted by a DM particle with the mass about 1.5 TeV if the nearby DM sub-halo locates at $\rm 0.1 \sim 0.3$ kpc away form the solar system. We fit the AMS-02 and DAMPE data assuming the DM annihilate into leptons with the branching ratio $e:\mu:\tau=1:1:1$. Such a condition can evade the constraints from CMB and the diffuse gamma rays from dwarf spheroidal galaxies (dSphs)~\cite{yuan2017}. In the fitting, we used numerical codes are \textsf{GALPROP}~\cite{galprop} and \textsf{DRAGON}~\cite{dragon} to calculate the propagation of CR electrons/positrons in the galaxy. We use the analytical solution presented in Ref.~\cite{Atoyan1995} to calculate the propagation of nearby CR electrons. In the first step, we use the \textsf{LikeDM} package~\cite{likedm} to calculate the likelihood (or $\chi^2$) and fit the AMS-02 and DAMPE data with power-lower background and extra astronomy contribution (see \cite{zu2017} for more details). Then we add the contribution of local DM halos directly as the local CR source only contributes the region around $\rm 1.5 TeV$. The fitting result is shown in Fig.~\ref{dampe}, in which the mass of DM particles is assumed as 1.5 TeV with the annihilation cross section $\langle \sigma v \rangle \simeq 3\times 10^{-26} cm^3 s^{-1}$ and the mass of nearby subhalo is $1 \rm \times 10^{8} m_\odot$ with a distance 0.1 kpc away from the solar system.

In order to satisfy DM annihilation cross section, $ \langle \sigma v \rangle \simeq 3\times 10^{-26} cm^3 s^{-1}$, required by DAMPE data, we focus on $P \otimes S$, $P \otimes P$, $V \otimes A$ and $V \otimes V$ couplings which can produce $s$-wave contributions in our following study.
In the following calculations, we assume a universal coupling of the mediator and three generation leptons, $g_\ell=g_e=g_\mu=g_\tau$. We implement our leptophilic DM model by using \textsf{FeynRules}~\cite{feynrule} and evaluate the DM relic density and annihilation cross section with \textsf{MicrOMEGAs}~\cite{micromega}. Since the mediators can induce the process $e^+e^- \to \ell^+\ell^-$, they are strongly constrained by LEP measurements of four-lepton contact interactions \cite{lep2} and di-lepton resonance searches in $e^+e^- \to \ell^+\ell^-\gamma$ \cite{Abbiendi:1999wm}. According the analysis in Ref.~\cite{Freitas:2014pua}, one can derive the following bounds of the coupling and mass of mediators $\Phi_{0,1}$ at 90\% C.L.,
\begin{eqnarray}
g^V_\ell /m_{\Phi_1} <
\begin{cases}
2.0\times 10^{-4} {\rm GeV}^{-1}, &m_{Z^\prime}>200 {\rm~GeV} \cr 6.9\times 10^{-4} {\rm~GeV}^{-1}, &100 {\rm~GeV}< m_{Z^\prime} <200 {\rm~GeV}
\end{cases}\\
g^A_\ell /m_{\Phi_1} <
\begin{cases}
2.4\times 10^{-4} {\rm GeV}^{-1}, &m_{Z^\prime}>200 {\rm~GeV} \cr 6.9\times 10^{-4} {\rm~GeV}^{-1}, &100 {\rm~GeV}< m_{Z^\prime} <200 {\rm~GeV}
\end{cases}\\
g^{S,P}_\ell /m_{\Phi_0} <
\begin{cases}
2.7\times 10^{-4} {\rm GeV}^{-1}, &m_{Z^\prime}>200 {\rm~GeV} \cr 7.3\times 10^{-4} {\rm~GeV}^{-1}, &100 {\rm~GeV}< m_{Z^\prime} <200 {\rm~GeV}
\end{cases}
\end{eqnarray}

\begin{figure}[ht]
  \centering
  \includegraphics[width=3in]{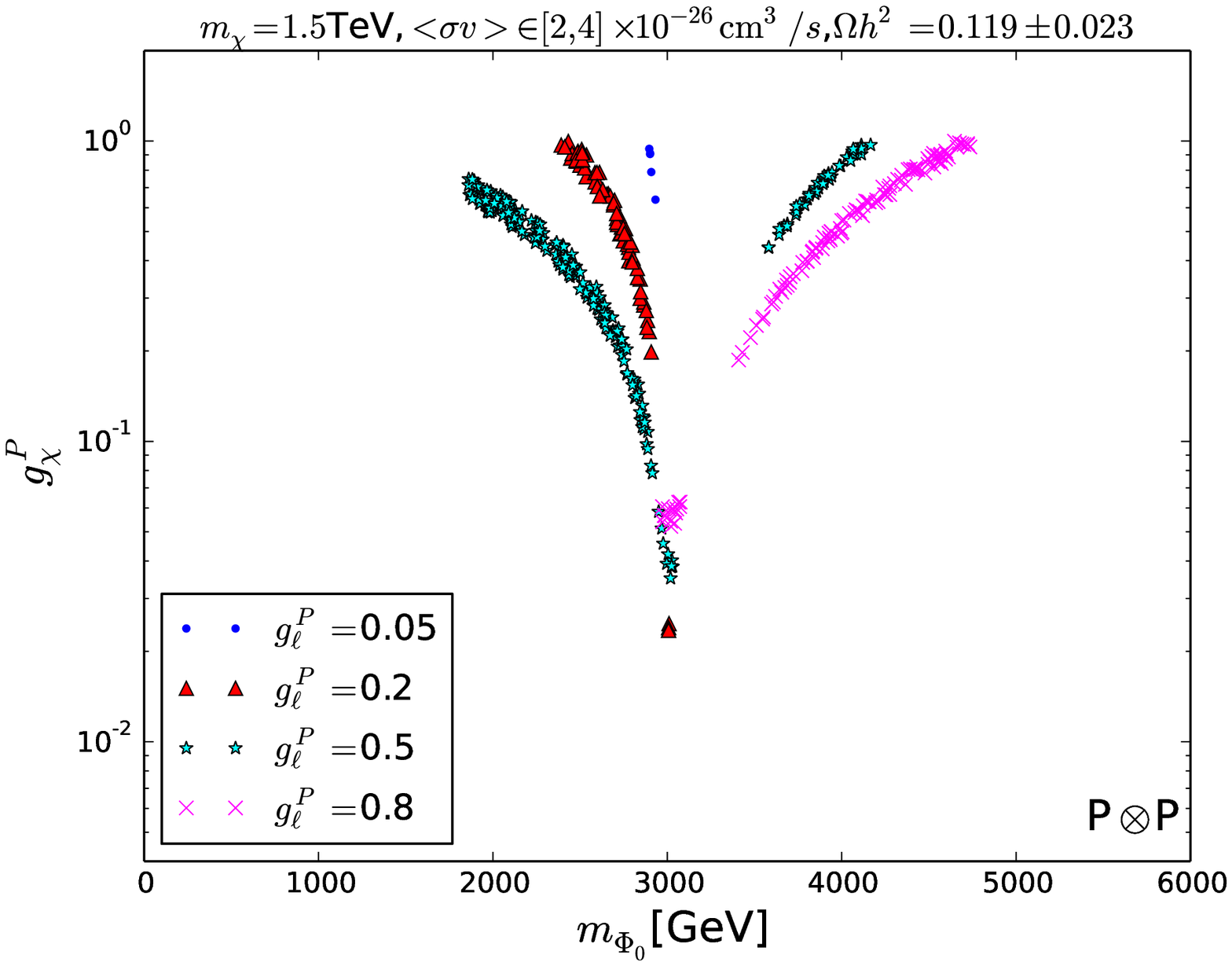}
  \includegraphics[width=3in]{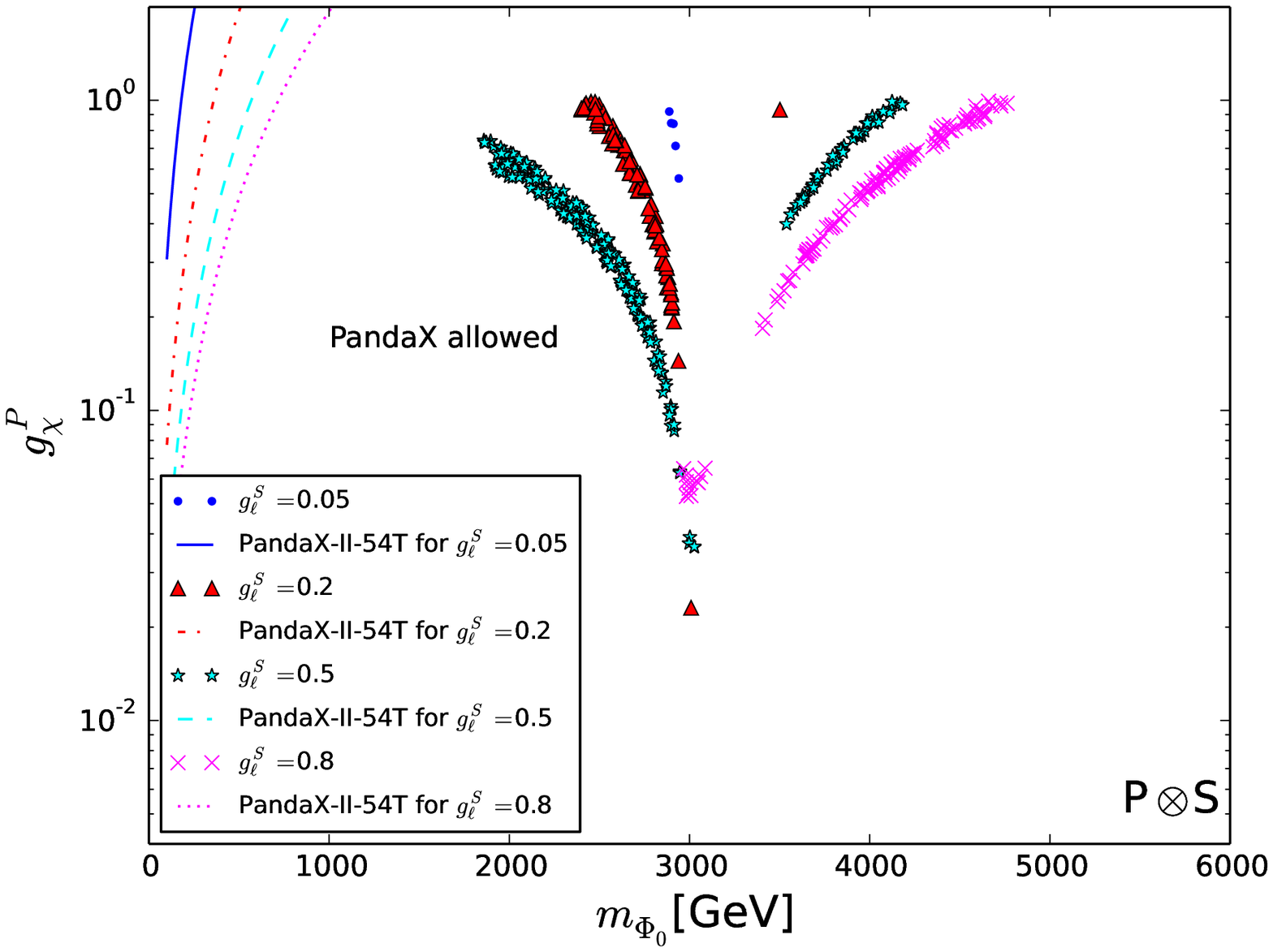}
  \includegraphics[width=3in]{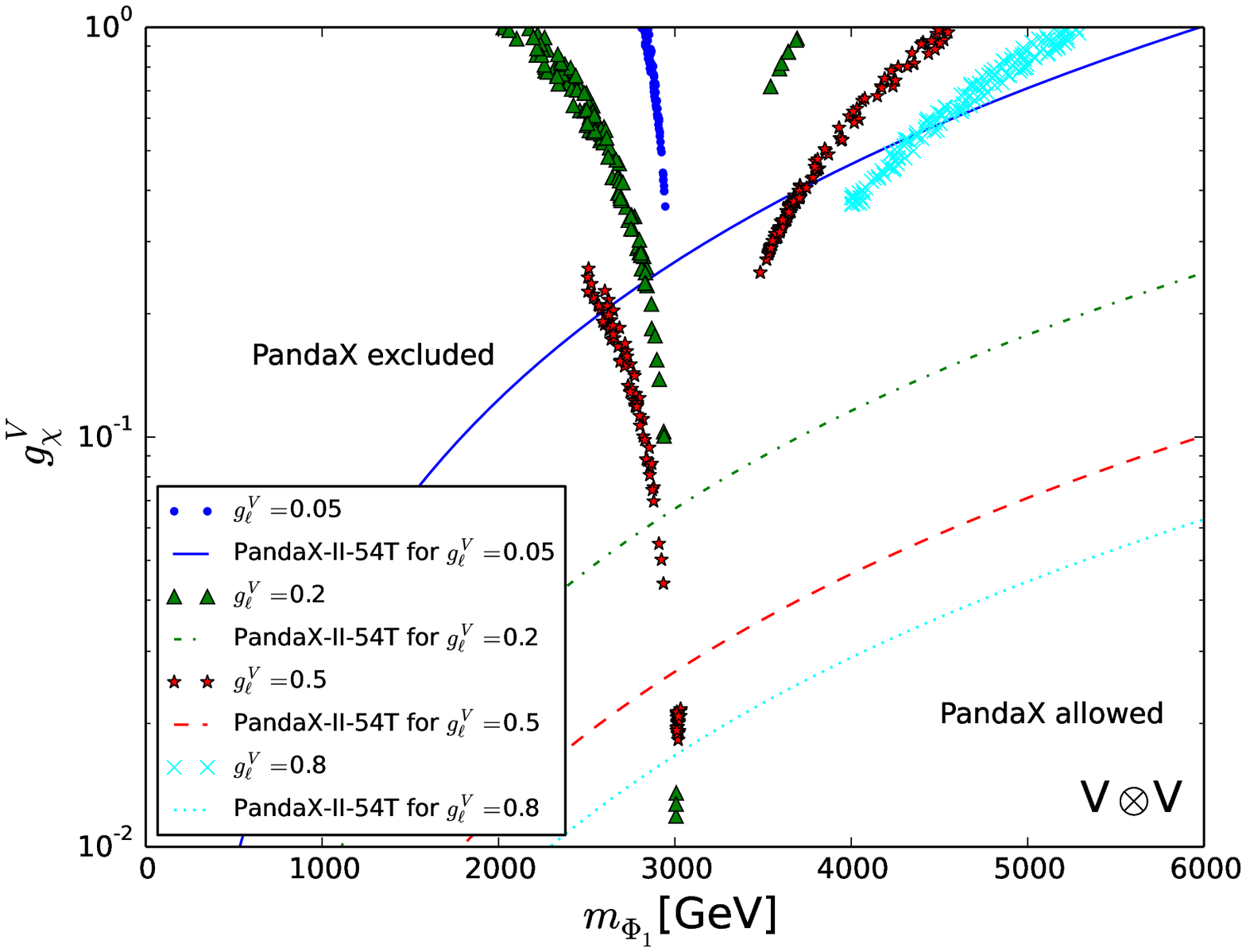}
  \includegraphics[width=3in]{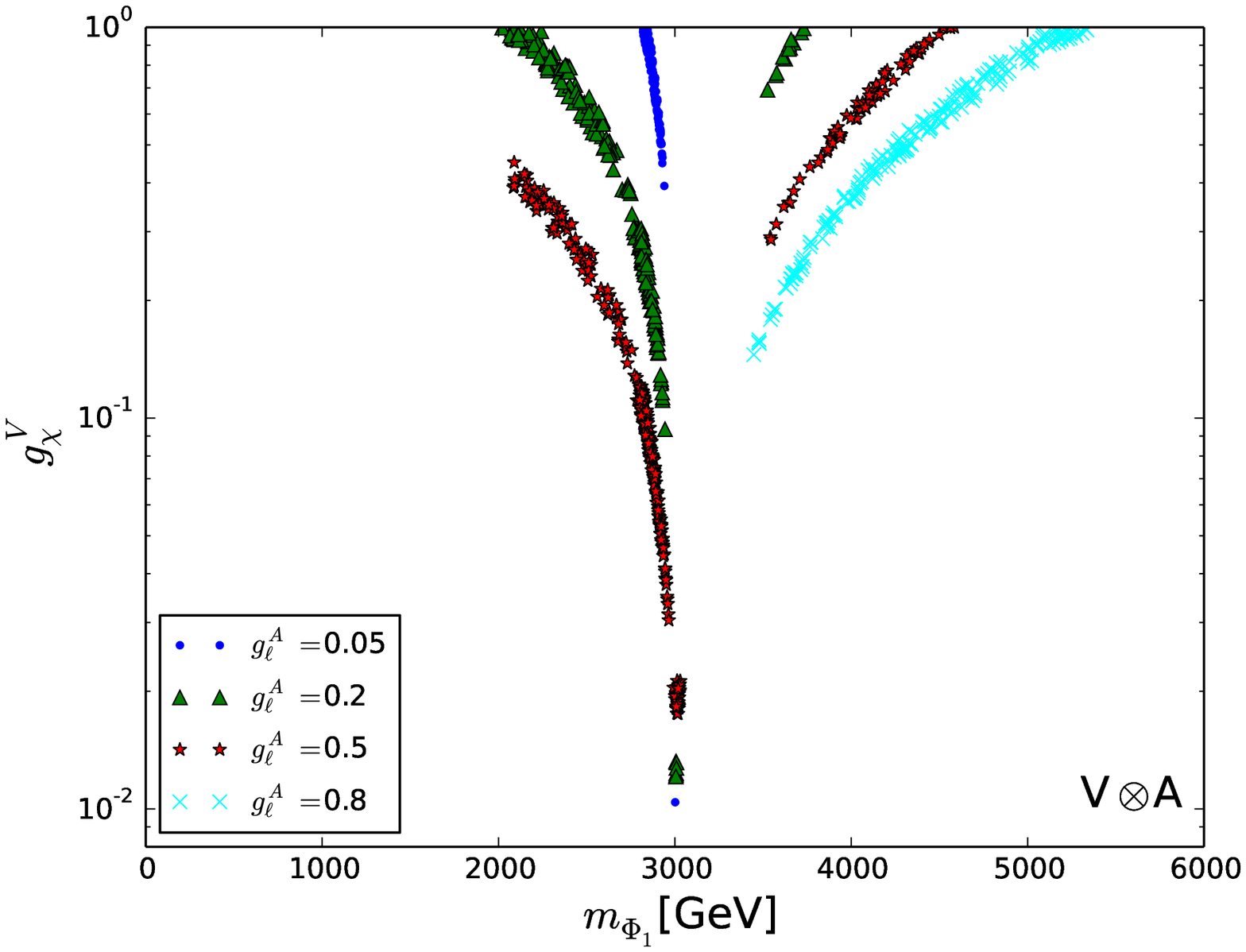}
\caption{Scatter plots of the samples satisfying the DM relic density within $2\sigma$ range of
Planck observed value, LEP bound and the DAMPE excess, projected on the plane of $g_\chi$ versus $m_\Phi$
for $g_\ell=0.05,02,0.5,0.8$. All samples are required to produce averaged annihilation cross-section $\langle \sigma v \rangle$ today within $ (2-4) \times 10^{-26} cm^3/s$. The 90\% C.L. exclusion limits from the current PandaX-II data is also shown~\cite{Pandax}. }
\label{nucleus}
\end{figure}
In Fig.~\ref{nucleus} we project the samples satisfying the requirements of DM relic density
within $2\sigma$ range of Planck observed value, LEP bound and the DAMPE excess on the plane of $g_\chi$
versus $m_\Phi$ for different values of $g_\ell$. All samples are required to produce averaged annihilation cross-section $\langle \sigma v \rangle$ today within $ (2-4) \times 10^{-26} cm^3/s$. When the mass of DM is close to $m_\Phi /2$,
DM annihilation cross section will be enhanced by resonance effect. In order to satisfy the DM
relic density requirement, the couplings $g^V_\chi$ and $g^V_\ell$ have to become small, which will
suppress the DM-nucleus scattering cross section so that the PandaX-II bound can be
evaded~\cite{Pandax}. For $P\otimes S$ coupling, the DM-nucleus scattering cross section is highly
reduced due to two-loop suppression, while for $P \otimes P$ and $V \otimes A$ couplings, the DM
has no interactions with nucleus. The surviving samples for $V \otimes V$ coupling are largely
excluded by the PandaX-II limits of DM-nucleus scattering. There are also limits from other direct detection
experiments such as XENON1T~\cite{Xenon1T} and LUX~\cite{LUX2016}. However, their current bounds are weaker than that of PandaX-II.



\begin{figure}[ht]
  \centering
   \includegraphics[width=3in]{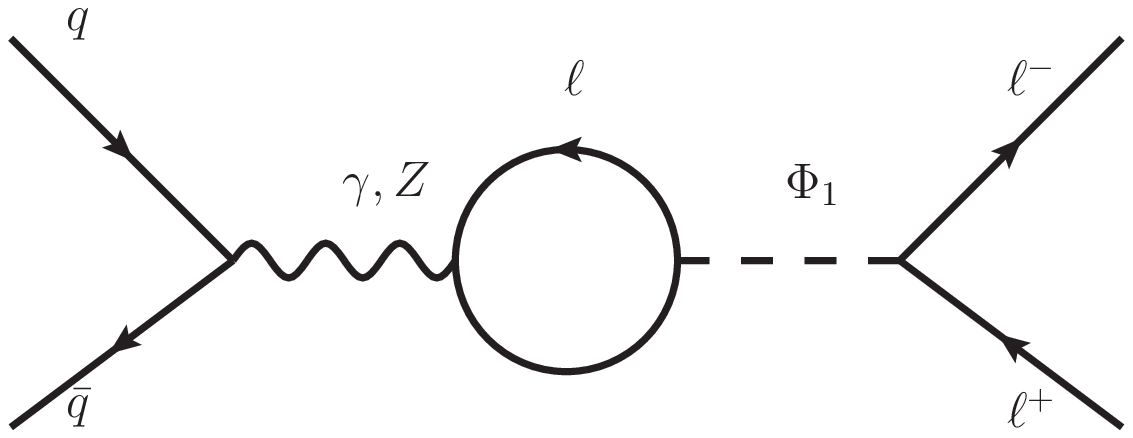}
 \caption{Feynman diagrams for the Drell-Yan process induced by a vector mediator $\Phi_1$ at the LHC.}
\label{feyn-dy}
\end{figure}

\begin{table}[ht]
\caption{The cross section of dilepton production $pp \to \Phi_1 \to \ell^+\ell^-$ at 13 TeV LHC, where the cross sections are in unit of fb. The benchmark points satisfy the DM relic density, the DAMPE $e^+ + e^-$ flux excess and the PandaX limits.}
\begin{tabular}{|c|c|c|c|}
  \hline
  & $(g_{\chi}^V,g_{\ell}^V)=(0.21,0.05)$ &$(g_{\chi}^V,g_{\ell}^V)=(0.012,0.2)$&$(g_{\chi}^V,g_{\ell}^V)=(0.2,0.5)$ \\
\hline
\hline
  $m_{\Phi_1}$=3TeV & $5.2\times10^{-4}$ & $7.5\times10^{-3}$ &$5.0 \times10^{-2}$\\
  \hline
\end{tabular}
\label{DY}
\end{table}


It should be mentioned that the vector mediator $\Phi_1$ can be produced at the LHC because of the loop-induced coupling between the mediator and light quarks, as shown in Fig.~\ref{feyn-dy}. The cross section in the narrow width limit is given by~\cite{DEramo:2017zqw}
\be
\sigma_{p p \rightarrow l^+ l^-} = \frac{\pi \, BR_{\Phi_1 \rightarrow l^+ l^-}}{3 s}
\sum_{q} C_{q\bar{q}} (m_{\Phi_1}^2 / s) \, \left({g^{V}_{q}}^2 + {g^{A}_{q}}^2\right) \ ,
\label{eq:sxLHCschannelNarrow}
\ee
where $BR_{\Phi_1 \rightarrow l^+ l^-}$ is the branching ratio of the decay $\Phi_1 \rightarrow l^+ l^-$. The parton luminosity $C_{q\bar{q}} (m_{\Phi_1}^2 / s)$ for the quark $q$ reads
\be
C_{q\bar{q}}(y) = \int_{y}^1 d x \, \frac{f_{q}(x) \, f_{\bar{q}}(y / x) + f_{q}(y / x) \, f_{\bar{q}}(x)}{x}  \ ,
\ee
with $f_{q, \bar{q}}(x)$ being the quark and antiquark parton distribution function (PDF).
We use MRST~\cite{mrst} to calculate the PDFs. The loop-induced couplings to quarks $g_q^V$ and $g_q^A$ are evaluated with package \textsf{runDM}~\cite{DEramo:2016gos}. The renormalization scale of the PDF and the couplings to quarks is set at $m_{\Phi_1}$. We choose some benchmark points that satisfy the DM relic density, the DAMPE $e^+ + e^-$ flux excess and the PandaX limits and calculate the corresponding
cross section of the dilepton process $pp \to \Phi_1 \to \ell^+\ell^-$ at the 13 TeV LHC, as given
in Table~\ref{DY}. We note that they are much lower than the current LHC-13 TeV sensitivity~\cite{Aaboud:2017buh}. We also evaluate the associated production processes $pp \to \ell^+\ell^- \Phi_{0,1}$ and
find they are negligibly small.

\begin{table}[ht]
\caption{Same as Table~\ref{DY}, but for the corrections to the anomalous magnetic moment of
the muon $\Delta a_\mu$.}
\begin{tabular}{|c|c|c|c|c|}
  \hline
  $m_{\Phi}$(TeV) & $(g_{\chi}^V,g_{\ell}^V)=(0.012,0.2)$ &$ (g_{\chi}^V,g_{\ell}^A)=(0.6,0.2)$& $(g_{\chi}^P,g_{\ell}^P)=(0.6,0.2)$& $(g_{\chi}^P,g_{\ell}^S)=(0.6,0.2)$ \\
\hline
\hline
  2.1 & --- & $-4.27\times10^{-12}$ &$-2.31\times10^{-11}$&$2.31\times10^{-11}$\\
  2.5 & --- &$-2.78\times10^{-12}$&$-1.65\times10^{-11}$&$1.65\times10^{-11}$  \\
  3.0 &$4.13\times10^{-13}$ & $-2\times10^{-12}$ &$-1.14\times10^{-11}$&$1.14\times10^{-11}$ \\
  3.6 &--- & $-1.43\times10^{-12}$&$-8.36\times10^{-12}$& $8.36\times10^{-12}$ \\
  \hline
\end{tabular}
\label{g-2}
\end{table}
In Table~\ref{g-2}, we give the corrections to the muon $g-2$ that arise from our leptophilic interactions~\cite{muong-2}. It can be seen that the couplings $V \otimes V$ and $P \otimes S$ can produce a positive correction, which, however, is less than the value
required by explaining the deviation of the muon $g-2$ from its experimental measurement.

\section{An $U(1)_X$ Realization }\label{section4}
An an example of realizations of LDM, we introduce a Dirac fermionic DM field ($\chi$) by imposing a $Z_2$ symmetry, under which all SM
matter particles are even while $\chi$ is odd. Besides, we add a new $U(1)_X$ gauge interactions for leptons only, with the corresponding gauge quantum numbers shown in Table~\ref{model}. The complex scalars $S$ and $T$ are introduced to break the $U(1)_X$ and $U(1)^\prime$ gauge symmetry, respectively.

\begin{table}[ht]
\label{model}
\centering
\begin{tabular}{|c|c|c|c|c|c|c|c|c|c|c|c|}
\hline
&$Q_L^a$ &$U_R^a$&$D_R^a$&$L_L^a$& $e_R$&~$\mu_R$&$\tau_R$& $S$&$T$&$F$&$\chi$\\
\hline
$U(1)_X$&0&0&0&~1&~1&~1&~1&$Q_S^X$&0&$Q_F^X$&$~0$\\
\hline
$U(1)^\prime$&0&0&0&~0&~0&~0&~1&$~0$&$Q_T^\prime$ &$Q_F^\prime$&$~Q_\chi^\prime$\\
\hline
\end{tabular}
 \caption{The $U(1)_X$ quantum number for SM matter contents with the generation index $a=(1,2,3)$. The complex scalars $S$ and $T$ are introduced to break the $U(1)_X$ and $U(1)^\prime$ gauge symmetry, respectively. The Dirac fermion $F$ that has charges of  $U(1)$, is introduced to generate kinetic mixing between the two $U(1)$ gauge bosons.}
\end{table}

Such an assignment will cause the $U(1)_X$ to be anomalous if there are no additional chiral fermions that are charged under $U(1)_X$ other than the SM matter contents. One possible way is to introduce new matter particles to cancel the anomaly. For example, we can add the fourth chiral-like family with non-trivial $U(1)_X$ quantum number, which satisfies the anomaly cancelation condition
 \beqa
 \sum\limits_i (3 n_i+m_i)+3k+l=0,
 \eeqa
with $n_i,m_i,k,l$ being the $U(1)_X$ quantum numbers for quarks($n_i$), leptons($m_i$) of the first
three family and the fourth family quarks($k$) and leptons ($l$), respectively, such as $l=-3m$ with
universal $m_i\equiv m$ for $e,\mu,\tau$ leptons and trivial quantum numbers for all quarks.
The fourth family can be very heavy by mixing with heavy vector-like fermions and can be compatible
with current collider constraints.

Since the DM direct detection experiments will give stringent constraints, we require that the Dirac fermion DM $\chi$ will not carry $U(1)_X$ quantum number but will transform non-trivially under an additional $U(1)^\pr$ gauge symmetry. Such $U(1)^\pr$ gauge symmetry will be broken by additional complex scalar field $T$. The couplings between DM and lepton pairs will be induced through kinetic mixing between $U(1)_X$ and $U(1)^\pr$. Given the gauge interaction $U(1)_X$ is universal for all kinds of leptons, we can anticipate that the decay products will lead to equal final states lepton species. This approach is similar to vector-portal DM scenario. Since the DM is vector-like, there will be no additional anomaly in the model.  New scalar $T$ or vector-like fermion $F$, which transform non-trivially under both $U(1)_X$ and $U(1)^\pr$, will induce non-trivial mixing between the two new $U(1)$ gauge symmetry through the following interactions,
\beqa
{\cal L}\supseteq |D_\mu S|^2+|D_\mu T|^2-m_S^2 |S|^2-m_T^2|T|^2-\la_1|S|^4-\la_2|T|^4 \nn \\ -\la_3|S|^2|T|^2+i\bar{F}\gamma^\mu D_\mu F-m_F\bar{F}F~,
\eeqa
 with
 \beqa
 D_\mu F&=&(\pa_\mu -i Q_F^X g_X A_\mu^X-i Q_F^\pr g^\pr A_\mu^\pr)F~,\nn\\
 D_\mu S&=&(\pa_\mu -i Q_S^X g_X A_\mu^X)S~,\nn\\
 D_\mu T&=&(\pa_\mu -i Q_T^\pr g^\pr A_\mu^\pr)T.
 \eeqa
As mentioned above, an odd $Z_2$ parity is imposed for the Dirac fermion $\chi$ to act as a viable DM candidate. The masses of the scalar $T$ are assumed to be heavier than the DM mass so that the DM will not annihilate into them. We should note that in the scalar potential, possible terms involving standard model Higgs fields $H$ as $(T^\da T)(H^\da H),(S^\da S)(H^\da H)$ etc could appear. Such terms could contribute to the DM direct detection at two loop level.

The kinetic mixing between two gauge bosons can be parameterized as
\beqa
{\cal L}\supseteq -\f{1}{4}F_{\mu\nu}F^{\mu\nu}-\f{1}{4}F^\pr_{\mu\nu}F^{\mu\nu\pr}-\f{\epsilon}{2}F^\pr_{\mu\nu}F^{\mu\nu}-\f{1}{2}m_1^2A_\mu A^\mu-\f{1}{2}m_2^2 A^{\mu\pr}A_{\mu\pr}~,
\eeqa
with
\beqa
\epsilon=-\f{g_X g^\pr }{12\pi^2}Q^X_F Q_F^\pr\log\(\f{m_F^2}{\mu^2}\)~,
\eeqa
after integrating out heavy fermion loops, or
\beqa
\epsilon=\f{g_1 g_2}{48\pi^2}Q_F Q_F^\pr\log\(\f{m_S^2}{\mu^2}\)~,
\eeqa
after integrating out possible heavy scalar loops.

The matrix to remove the mixing is given as
\beqa
\(\bea{c}\tl{A}_\mu\\\tl{A}^\pr_\mu\eea\)=\(\bea{cc}\f{1}{\sqrt{1+\epsilon^2}}& 0\\-\f{\epsilon}{\sqrt{1+\epsilon^2}}&1 \eea\)\(\bea{c}{A}_\mu\\{A}^\pr_\mu\eea\),
\eeqa
with the Lagrangian involving the mass mixing
\beqa
{\cal L}=-\f{1}{4}\tl{F}_{\mu\nu}\tl{F}^{\mu\nu}-\f{1}{4}\tl{F}^\pr_{\mu\nu}\tl{F}^{\mu\nu\pr}-\f{1}{2}m_1^2\tl{A}_\mu \tl{A}^\mu-\f{1}{2}m_2^2
\tl{A}^\pr_\mu \tl{A}^{\mu\pr}-m_1^2\epsilon \tl{A}_\mu \tl{A}^{\mu\pr}~.
\eeqa

Assuming identical masses for the scalars $m_1^2=m_2^2$, we obtain
\beqa
\(\bar{\chi}\gamma^\mu\chi\) \(\bar{L}\gamma^\nu L\) \[\epsilon/m_2^2\].
\eeqa

To explain the DAMPE excess without conflicting with direct detection experiments, we can choose $m_2\simeq 3$ TeV and the mixing parameter $\epsilon\approx 1.0\tm 10^{-2}$. Such values can be obtained by requiring $g_1=g_2\approx 0.3$ with $Q_1=Q_2=1$.

\section{Conclusion}\label{section5}
In this work, we explained the recent DAMPE cosmic $e^+ + e^-$ excess in simplified leptophilic Dirac
fermion dark matter (LDM) framework with a scalar ($\Phi_0$) or vector ($\Phi_1$) mediator. We found
that the couplings $P \otimes S$, $P \otimes P$, $V \otimes A$ and $V \otimes V$ can fit the DAMPE
data under the constraints from gamma-rays and cosmic-rays. However,
for the $V \otimes V$ coupling, due to the stringent constraints from the PandaX-II data,
the surviving samples only exist in the resonance region, $m_{\Phi_1} \simeq 2m_{\chi}$.
But for other couplings, the direct detection bounds can easily be evaded. We also studied the
possible collider signatures of LDM, such as the Drell-Yan process $pp \to \Phi_1 \to \ell^+\ell^-$,
and the muon $g-2$. In the end, we constructed an $U(1)$ extension of the SM to realize our
simplified LDM model.

\section*{Acknowledgement}
G. Duan was supported by a visitor program of Nanjing Normal University, during which this work
was finished. This work was supported by the National Natural Science Foundation of China (NNSFC)
under grant No. 11705093, 11675242 and 11773075, by the CAS Center for Excellence in Particle
Physics (CCEPP),  by the CAS Key Research Program of Frontier Sciences and by a Key R\&D Program
of Ministry of Science and Technology of China under number 2017YFA0402200-04.
FL is also supported by the Youth Innovation Promotion Association of
Chinese Academy of Sciences (No. 2016288).


\begin{thebibliography}{99}

\bibitem{dampe1} J. Chang, Chinese Journal of Space Science 34, 550 (2014).

\bibitem{dampe2}
  J.~Chang {\it et al.} [DAMPE Collaboration],
  Astropart.\ Phys.\  {\bf 95}, 6 (2017)
  doi:10.1016/j.astropartphys.2017.08.005
  [arXiv:1706.08453 [astro-ph.IM]].

\bibitem{dampe-data}
  G.~Ambrosi {\it et al.} [DAMPE Collaboration],
  doi:10.1038/nature24475
  arXiv:1711.10981 [astro-ph.HE].

\bibitem{yuan2017}
  Q.~Yuan {\it et al.},
  arXiv:1711.10989 [astro-ph.HE].

\bibitem{bi}
  K.~Fang, X.~J.~Bi and P.~F.~Yin,
  arXiv:1711.10996 [astro-ph.HE].

\bibitem{Fan:2017sor}
  Y.~Z.~Fan, W.~C.~Huang, M.~Spinrath, Y.~L.~S.~Tsai and Q.~Yuan,
  arXiv:1711.10995 [hep-ph].

\bibitem{Gu:2017gle}
  P.~H.~Gu and X.~G.~He,
  arXiv:1711.11000 [hep-ph].


\bibitem{ldm-1}
  S.~Chang, R.~Edezhath, J.~Hutchinson and M.~Luty,
  Phys.\ Rev.\ D {\bf 90}, no. 1, 015011 (2014)
  doi:10.1103/PhysRevD.90.015011
  [arXiv:1402.7358 [hep-ph]].

\bibitem{ldm-2}
  D.~Schmidt, T.~Schwetz and T.~Toma,
  Phys.\ Rev.\ D {\bf 85}, 073009 (2012)
  doi:10.1103/PhysRevD.85.073009
  [arXiv:1201.0906 [hep-ph]].

\bibitem{ldm-3}
  P.~Agrawal, S.~Blanchet, Z.~Chacko and C.~Kilic,
  Phys.\ Rev.\ D {\bf 86}, 055002 (2012)
  doi:10.1103/PhysRevD.86.055002
  [arXiv:1109.3516 [hep-ph]].

\bibitem{ldm-4}
  C.~D.~Carone and R.~Primulando,
  Phys.\ Rev.\ D {\bf 84}, 035002 (2011)
  doi:10.1103/PhysRevD.84.035002
  [arXiv:1105.4635 [hep-ph]].

\bibitem{ldm-5}
  P.~Ko and Y.~Omura,
  Phys.\ Lett.\ B {\bf 701}, 363 (2011)
  doi:10.1016/j.physletb.2011.06.009
  [arXiv:1012.4679 [hep-ph]].

\bibitem{ldm-6}
  N.~Haba, Y.~Kajiyama, S.~Matsumoto, H.~Okada and K.~Yoshioka,
  Phys.\ Lett.\ B {\bf 695}, 476 (2011)
  doi:10.1016/j.physletb.2010.11.063
  [arXiv:1008.4777 [hep-ph]].

\bibitem{ldm-7}
  Y.~Farzan, S.~Pascoli and M.~A.~Schmidt,
  JHEP {\bf 1010}, 111 (2010)
  doi:10.1007/JHEP10(2010)111
  [arXiv:1005.5323 [hep-ph]].


\bibitem{ldm-8}
  E.~J.~Chun, J.~C.~Park and S.~Scopel,
  JCAP {\bf 1002}, 015 (2010)
  doi:10.1088/1475-7516/2010/02/015
  [arXiv:0911.5273 [hep-ph]].

\bibitem{ldm-9}
  T.~Cohen and K.~M.~Zurek,
  Phys.\ Rev.\ Lett.\  {\bf 104}, 101301 (2010)
  doi:10.1103/PhysRevLett.104.101301
  [arXiv:0909.2035 [hep-ph]].

\bibitem{ldm-10}
  H.~Davoudiasl,
  Phys.\ Rev.\ D {\bf 80}, 043502 (2009)
  doi:10.1103/PhysRevD.80.043502
  [arXiv:0904.3103 [hep-ph]].

\bibitem{ldm-11}
  A.~Ibarra, A.~Ringwald, D.~Tran and C.~Weniger,
  JCAP {\bf 0908}, 017 (2009)
  doi:10.1088/1475-7516/2009/08/017
  [arXiv:0903.3625 [hep-ph]].

\bibitem{ldm-12}
  B.~Kyae,
  JCAP {\bf 0907}, 028 (2009)
  doi:10.1088/1475-7516/2009/07/028
  [arXiv:0902.0071 [hep-ph]].

\bibitem{ldm-13}
  C.~R.~Chen and F.~Takahashi,
  JCAP {\bf 0902}, 004 (2009)
  doi:10.1088/1475-7516/2009/02/004
  [arXiv:0810.4110 [hep-ph]].

\bibitem{ldm-14}
  E.~A.~Baltz and L.~Bergstrom,
  Phys.\ Rev.\ D {\bf 67}, 043516 (2003)
  doi:10.1103/PhysRevD.67.043516
  [hep-ph/0211325].

\bibitem{ldm-15}
  Y.~Bai and J.~Berger,
  JHEP {\bf 1408}, 153 (2014)
  doi:10.1007/JHEP08(2014)153
  [arXiv:1402.6696 [hep-ph]].

\bibitem{ldm-16}
  P.~Schwaller, T.~M.~P.~Tait and R.~Vega-Morales,
  Phys.\ Rev.\ D {\bf 88}, no. 3, 035001 (2013)
  doi:10.1103/PhysRevD.88.035001
  [arXiv:1305.1108 [hep-ph]].

\bibitem{ldm-17}
  L.~Basso, O.~Fischer and J.~J.~van der Bij,
  Phys.\ Rev.\ D {\bf 87}, no. 3, 035015 (2013)
  doi:10.1103/PhysRevD.87.035015
  [arXiv:1207.3250 [hep-ph]].

\bibitem{ldm-18}
  C.~D.~Carone, A.~Cukierman and R.~Primulando,
  Phys.\ Lett.\ B {\bf 704}, 541 (2011)
  doi:10.1016/j.physletb.2011.09.086
  [arXiv:1108.2084 [hep-ph]].

\bibitem{ldm-19}
  W.~Chao,
  Phys.\ Lett.\ B {\bf 695}, 157 (2011)
  doi:10.1016/j.physletb.2010.10.056
  [arXiv:1005.1024 [hep-ph]].


\bibitem{ldm-20}
  S.~Khalil, H.~S.~Lee and E.~Ma,
  Phys.\ Rev.\ D {\bf 79}, 041701 (2009)
  doi:10.1103/PhysRevD.79.041701
  [arXiv:0901.0981 [hep-ph]].


\bibitem{ldm-21}
  Q.~H.~Cao, E.~Ma and G.~Shaughnessy,
  Phys.\ Lett.\ B {\bf 673}, 152 (2009)
  doi:10.1016/j.physletb.2009.02.015
  [arXiv:0901.1334 [hep-ph]].

\bibitem{ldm-22}
  A.~Freitas and S.~Westhoff,
  JHEP {\bf 1410}, 116 (2014)
  doi:10.1007/JHEP10(2014)116
  [arXiv:1408.1959 [hep-ph]].

\bibitem{ldm-23}
  N.~F.~Bell, Y.~Cai, R.~K.~Leane and A.~D.~Medina,
  Phys.\ Rev.\ D {\bf 90}, no. 3, 035027 (2014)
  [arXiv:1407.3001 [hep-ph]].


\bibitem{ldm-24}
  M.~C.~Chen, J.~Huang and V.~Takhistov,
  JHEP {\bf 1602}, 060 (2016)
  doi:10.1007/JHEP02(2016)060
  [arXiv:1510.04694 [hep-ph]].

\bibitem{ldm-25}
  J.~Kile, A.~Kobach and A.~Soni,
  Phys.\ Lett.\ B {\bf 744}, 330 (2015)
  doi:10.1016/j.physletb.2015.04.005
  [arXiv:1411.1407 [hep-ph]].

\bibitem{ldm-26}
  S.~Chang, R.~Edezhath, J.~Hutchinson and M.~Luty,
  Phys.\ Rev.\ D {\bf 90}, no. 1, 015011 (2014)
  doi:10.1103/PhysRevD.90.015011
  [arXiv:1402.7358 [hep-ph]].

\bibitem{ldm-27}
  J.~Kopp, L.~Michaels and J.~Smirnov,
  JCAP {\bf 1404}, 022 (2014)
  doi:10.1088/1475-7516/2014/04/022
  [arXiv:1401.6457 [hep-ph]].

\bibitem{ldm-28}
  K.~Belotsky, M.~Khlopov, C.~Kouvaris and M.~Laletin,
  Adv.\ High Energy Phys.\  {\bf 2014}, 214258 (2014)
  doi:10.1155/2014/214258
  [arXiv:1403.1212 [astro-ph.CO]].

\bibitem{ldm-29}
  P.~S.~B.~Dev, D.~K.~Ghosh, N.~Okada and I.~Saha,
  Phys.\ Rev.\ D {\bf 89}, 095001 (2014)
  doi:10.1103/PhysRevD.89.095001
  [arXiv:1307.6204 [hep-ph]].


\bibitem{ldm-30}
  A.~Alves, A.~Berlin, S.~Profumo and F.~S.~Queiroz,
  Phys.\ Rev.\ D {\bf 92}, no. 8, 083004 (2015)
  [arXiv:1501.03490 [hep-ph]].

\bibitem{ldm-31}
S.~M.~Boucenna, M.~Chianese, G.~Mangano, G.~Miele, S.~Morisi, O.~Pisanti and E.~Vitagliano,
JCAP {\bf 1512}, no. 12, 055 (2015)
[arXiv:1507.01000 [hep-ph]].



\bibitem{ldm-32}
A.~Berlin, D.~Hooper and S.~D.~McDermott,
Phys.\ Rev.\ D {\bf 89}, no. 11, 115022 (2014)
[arXiv:1404.0022 [hep-ph]].

\bibitem{ldm-33}
S.~Dutta, D.~Sachdeva and B.~Rawat,
Eur.\ Phys.\ J.\ C {\bf 77}, no. 9, 639 (2017)
[arXiv:1704.03994 [hep-ph]].

\bibitem{gldm-1}
  M.~Das and S.~Mohanty,
  Phys.\ Rev.\ D {\bf 89}, no. 2, 025004 (2014)
  doi:10.1103/PhysRevD.89.025004
  [arXiv:1306.4505 [hep-ph]].

\bibitem{gldm-2}
  P.~J.~Fox and E.~Poppitz,
  Phys.\ Rev.\ D {\bf 79}, 083528 (2009)
  doi:10.1103/PhysRevD.79.083528
  [arXiv:0811.0399 [hep-ph]].

\bibitem{gldm-3}
  X.~J.~Bi, X.~G.~He and Q.~Yuan,
  Phys.\ Lett.\ B {\bf 678}, 168 (2009)
  doi:10.1016/j.physletb.2009.06.009
  [arXiv:0903.0122 [hep-ph]].

\bibitem{gldm-4}
  J.~Kopp, L.~Michaels and J.~Smirnov,
  JCAP {\bf 1404}, 022 (2014)
  doi:10.1088/1475-7516/2014/04/022
  [arXiv:1401.6457 [hep-ph]].


\bibitem{gldm-5}
  A.~Hamze, C.~Kilic, J.~Koeller, C.~Trendafilova and J.~H.~Yu,
  Phys.\ Rev.\ D {\bf 91}, no. 3, 035009 (2015)
  doi:10.1103/PhysRevD.91.035009
  [arXiv:1410.3030 [hep-ph]].

\bibitem{gldm-6}
  C.~J.~Lee and J.~Tandean,
  JHEP {\bf 1504}, 174 (2015)
  doi:10.1007/JHEP04(2015)174
  [arXiv:1410.6803 [hep-ph]].

\bibitem{gldm-7}
  S.~Baek and P.~Ko,
  JCAP {\bf 0910}, 011 (2009)
  doi:10.1088/1475-7516/2009/10/011
  [arXiv:0811.1646 [hep-ph]].

\bibitem{gldm-8}
  F.~del Aguila, M.~Chala, J.~Santiago and Y.~Yamamoto,
  JHEP {\bf 1503}, 059 (2015)
  doi:10.1007/JHEP03(2015)059
  [arXiv:1411.7394 [hep-ph]].

\bibitem{gldm-9}
  B.~Fornal, Y.~Shirman, T.~M.~P.~Tait and J.~R.~West,
  Phys.\ Rev.\ D {\bf 96}, no. 3, 035001 (2017)
  doi:10.1103/PhysRevD.96.035001
  [arXiv:1703.00199 [hep-ph]].

\bibitem{gldm-10}
  B.~Fornal,
  Mod.\ Phys.\ Lett.\ A {\bf 32}, no. 19, 1730018 (2017)
  doi:10.1142/S021773231730018X
  [arXiv:1705.07297 [hep-ph]].


\bibitem{review-ldm}
  J.~Kile,
  Mod.\ Phys.\ Lett.\ A {\bf 28}, 1330031 (2013)
  doi:10.1142/S0217732313300310
  [arXiv:1308.0584 [hep-ph]].

\bibitem{Kopp:2009et}
J.~Kopp, V.~Niro, T.~Schwetz and J.~Zupan,
Phys.\ Rev.\ D {\bf 80}, 083502 (2009)
[arXiv:0907.3159 [hep-ph]].


\bibitem{galprop} A. W. Strong and I. V. Moskalenko, Astrophys. J. {\bf 509}, 212 (1998), astro-ph/9807150.

\bibitem{dragon} C. Evoli, D. Gaggero, D. Grasso and L. Maccione, J. Cosmol. Astropart. Phys. {\bf 10}, 018 (2008), 0807.4730.

\bibitem{Atoyan1995} A. M. Atoyan, F. A. Aharonian, and H. J. V¡§olk, Phys. Rev. D
52, 3265 (1995).


\bibitem{likedm} X. Huang, Y.-L. S. Tsai, and Q. Yuan, Comput. Phys. Commun. {\bf 213}, 252 (2017), 1603.07119.

\bibitem{zu2017} Lei Zu, Cun Zhang, Lei Feng, Qiang Yuan and Yi-Zhong Fan, arXiv:1711.11052.



\bibitem{feynrule}
A.~Alloul, N.~D.~Christensen, C.~Degrande, C.~Duhr and B.~Fuks,
Comput.\ Phys.\ Commun.\  {\bf 185}, 2250 (2014)
[arXiv:1310.1921 [hep-ph]].

\bibitem{micromega}
 G. Belanger et al., Comput. Phys. Commun. 182, 842 (2011)



\bibitem{lep2}
  S.~Schael {\it et al.}  [ALEPH and DELPHI and L3 and OPAL and LEP Electroweak Collaborations],
  Phys.\ Rept.\  {\bf 532}, 119 (2013)
  [arXiv:1302.3415 [hep-ex]].

\bibitem{Abbiendi:1999wm}
  G.~Abbiendi {\it et al.}  [OPAL Collaboration],
  Eur.\ Phys.\ J.\ C {\bf 13}, 553 (2000)
  [hep-ex/9908008].

\bibitem{Freitas:2014pua}
  A.~Freitas, J.~Lykken, S.~Kell and S.~Westhoff,
  JHEP {\bf 1405}, 145 (2014)
  [arXiv:1402.7065 [hep-ph]].

\bibitem{Pandax}
X.~Cui {\it et al.} [PandaX-II Collaboration],
Phys.\ Rev.\ Lett.\  {\bf 119}, no. 18, 181302 (2017)
[arXiv:1708.06917 [astro-ph.CO]].

\bibitem{Xenon1T}
E.~Aprile {\it et al.} [XENON Collaboration],
Phys.\ Rev.\ Lett.\  {\bf 119}, no. 18, 181301 (2017)
[arXiv:1705.06655 [astro-ph.CO]].

\bibitem{LUX2016}
  D.~S.~Akerib {\it et al.},
  arXiv:1608.07648 [astro-ph.CO].

\bibitem{DEramo:2017zqw}
F.~D'Eramo, B.~J.~Kavanagh and P.~Panci,
Phys.\ Lett.\ B {\bf 771}, 339 (2017)
[arXiv:1702.00016 [hep-ph]].


\bibitem{mrst}
  A.~D.~Martin, W.~J.~Stirling, R.~S.~Thorne and G.~Watt,
  Phys.\ Lett.\ B {\bf 652}, 292 (2007)
  doi:10.1016/j.physletb.2007.07.040
  [arXiv:0706.0459 [hep-ph]].



\bibitem{DEramo:2016gos}
F.~D'Eramo, B.~J.~Kavanagh and P.~Panci,
JHEP {\bf 1608}, 111 (2016)
[arXiv:1605.04917 [hep-ph]].



\bibitem{Aaboud:2017buh}
  M.~Aaboud {\it et al.} [ATLAS Collaboration],
JHEP {\bf 1710}, 182 (2017)
[arXiv:1707.02424 [hep-ex]].

\bibitem{muong-2}
P.~Agrawal, Z.~Chacko and C.~B.~Verhaaren,
JHEP {\bf 1408}, 147 (2014)
[arXiv:1402.7369 [hep-ph]].




\end{thebibliography}
\end{document}